# From PREVENTion to REACTion: Enhancing Failure Resolution in Naval Systems

Maria Teresa Rossi
*University of Milano-Bicocca*
Milano, Italy
maria.rossi@unimib.it

Leonardo Mariani
*University of Milano-Bicocca*
Milano, Italy
leonardo.mariani@unimib.it

Oliviero Riganelli
*University of Milano-Bicocca*
Milano, Italy
oliviero.riganelli@unimib.it

*Abstract*—Complex and large industrial systems often misbehave, for instance, due to wear, misuse, or faults. To cope with these incidents, it is important to timely detect their occurrences, localize the sources of the problems, and implement the appropriate countermeasures. This paper reports our experience with a state-of-the-art failure prediction method, PREVENT, and its extension with a troubleshooting module, REACT, applied to naval systems developed by Fincantieri. Our results show how to integrate anomaly detection with troubleshooting procedures. We conclude by discussing a lesson learned, which may help deploy and extend these analyses to other industrial products.

## I. INTRODUCTION

Many large industrial systems are composed of a mix of mechanical, hardware, and software components that may interact in unpredictable ways and cause failures. For example, components may stop working due to wear, misuse, and faults. Failure prediction and fault localization approaches help to deal with these problems as soon as they arise, timely reporting any misbehavior to users, and possibly localizing the root cause of the problem, respectively.

Some approaches address these challenges with rule-based methods [4], which rely on failure patterns defined by domain experts, or signature-based methods [8], [10], [13], [19], which depend on past failures to mine failure patterns. Both cases are limited by the need to know or have access to typical issues, which is often not the case in real industrial cases.

PREVENT [6] has recently been proposed as an unsupervised approach that integrates the capability to predict failures before they are observable by users and localize the responsible components. Although PREVENT has been validated with non-trivial systems, no studies have investigated whether it is effective in the context of industrial systems with a complex interplay of heterogeneous components. The *first contribution* of this paper is the empirical assessment of PREVENT with data obtained from naval systems operated by Fincantieri, our partner in a joint innovation project funded by the Italian PNRR program.

Further, while PREVENT can report anomalies and localize faulty components, it does not provide guidance on how to address them, leaving operators to manually search through extensive documentation, which can be a slow and error-prone task. The *second contribution* of this paper is an extension of PREVENT with REACT, a module that exploits Retrieval Augmented Generation (RAG)[1] to automatically extract troubleshooting procedures from manuals based on a set of anomalous indicators. In particular, REACT automatically generates a prompt from a set of anomalies returned by PREVENT. The prompt is designed to retrieve the right procedures to be applied to solve an ongoing problem, drastically accelerating the reaction time of the operator. The role of the RAG is important in bridging the gap between low-level system indicators and the language used in the manuals. Low-level indicators describe internal behaviors, while procedures refer to actions executable on the systems' interfaces. The flexibility of the RAG potentially allows for the retrieval of relevant procedures even when the terms used in the queries differ from those used in the documentation.

We report our experience with the application of PREVENT and REACT to a system operated by Fincantieri[2], namely the *Air Compressed System*, which plays a critical role in ensuring the safe and efficient operation of ships. To increase the generality of the reported evidence, we assessed PREVENT also with a *CNC milling machine* from a public dataset provided by the University of Michigan.

Our investigation shows that PREVENT must be tuned to avoid generating too many false positives, and that its integration with REACT can promisingly link low-level anomalies to user-oriented troubleshooting procedures.

The paper is organized as follows. Section II describes our extension of PREVENT with REACT. Section III outlines the methodology and presents the empirical results. Section IV distills a lesson learned. Section V discusses related work. Finally, Section VI provides final remarks.

## II. PROPOSED APPROACH

Figure 1 illustrates the interaction between the two main components of our approach: PREVENT and REACT.

The first component, PREVENT, operates on the *Key Performance Indicators (KPIs) continuously collected during system operation* to (a) detect anomalous KPIs, which are forwarded to REACT, and (b) identify the root cause KPIs, which are presented to the operator. Both anomalous and root cause KPIs are generated only when PREVENT predicts a failure.

---

[1]A RAG combines a knowledge base, populated with relevant documents, with an LLM, which is used to generate responses. A RAG automatically augments the prompts with semantically relevant information extracted from the knowledge base.

[2]https://www.fincantieri.com/

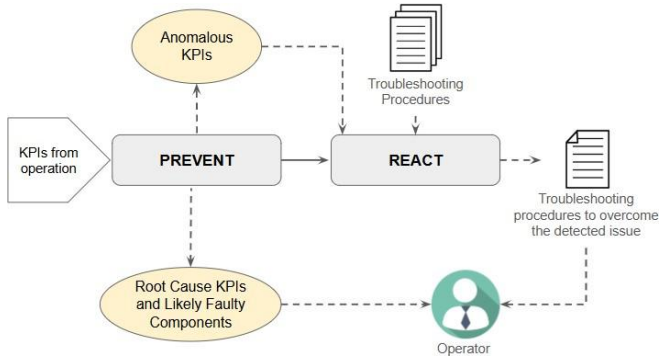

Fig. 1. Overview of the approach.

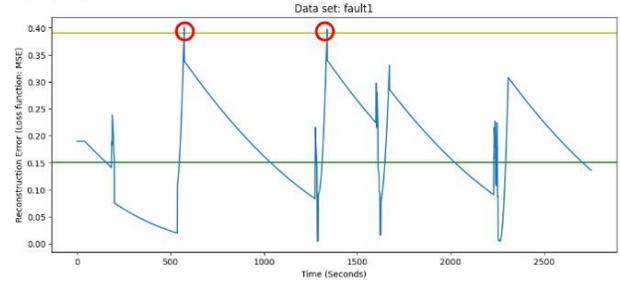

Fig. 3. Example of reconstruction error trend.

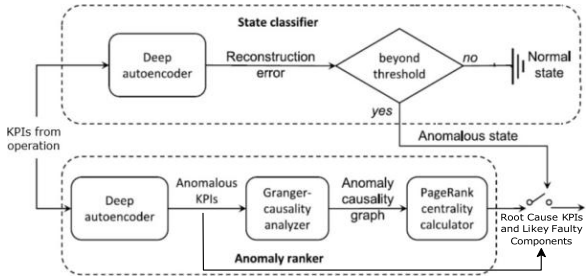

Fig. 2. Overview of PREVENT from [6].

Note that the *Anomalous KPIs* represent the full set of currently anomalous KPIs, giving a fairly complete overview of the issue affecting the system. For this reason, the cascade anomalies caused by other anomalies are part of this list. For example, a performance degradation in one component may propagate and trigger similar issues in dependent components, resulting in multiple anomalous KPIs. All such KPIs are passed to REACT, which aims at retrieving troubleshooting procedures that are capable of addressing the issues that occur across all the malfunctioning components. While PREVENT sends to the operators only the *root cause KPIs and the likely faulty components*, which jointly represent the likely root cause of the problem, so that operators can focus their attention on the most suspicious components only.

### A. Prevent

PREVENT [6] combines a deep autoencoder, Granger causality analysis, and PageRank centrality analysis to predict failures and localize faulty components without requiring labeled training data, as shown in Figure 2. The main components are a State Classifier and an Anomaly Ranker, which detect anomalous states and identify the root cause KPIs, respectively. These components process time series of Key Performance Indicators (KPIs), which are sets of metric values observed by monitoring an application at regular intervals. In particular, a KPI is a pair (metric, node), representing a metric value collected from either a digital or physical resource of the monitored application.

Both the State Classifier and the Anomaly Ranker include a deep autoencoder that can be trained with unsupervised training data containing KPI data collected during normal (non-failing) executions. The State Classifier detects anomalous states, which indicate that the observed system is not operating correctly. The Anomaly Ranker produces a list of root cause KPIs ordered by anomaly relevance and the likely faulty components, suggesting the possible sources of any problem that is affecting the system. To limit the generation of false positives, the root cause KPIs are output only if the current state is anomalous.

The State Classifier analyzes KPI values observed in production by comparing them to patterns learned from normal executions, in order to detect deviations from expected system behavior. Specifically, a state is flagged as anomalous if its reconstruction error exceeds the mean reconstruction error observed during training by more than a threshold, defined as the standard deviation multiplied by parameter *sigma*. Intuitively this happens when the current scenario largely differs from the training-time scenario. Figure 3 illustrates an example distribution of reconstruction error over time. The yellow line represents the threshold calculated by multiplying the standard deviation by *sigma*. When the values of the mean reconstruction error exceed this threshold (e.g., the two points circled in red in Figure 3), the state classifier labels the state as anomalous.

The Anomaly Ranker identifies KPIs that deviate significantly from the values observed during training and ranks anomalous nodes based on their relevance to these KPIs. Specifically, it constructs a graph representing the causality dependencies between the anomalous KPIs using Granger causality analysis [3]. Based on the causality graph, the ranker calculates the PageRank centrality [14] value for each anomalous KPI, distinguishing the anomalous KPIs that likely represent the root cause of the problem (i.e., the central KPIs according to the causal propagation of the anomalies) from the anomalous KPIs that are a side-effect of the problem (i.e., the non-central KPIs that result from the direct and indirect causal propagation of the root anomalous values). Each anomalous KPI is associated with an anomaly score that is the reconstruction error. Finally, the Anomaly ranker outputs the three components that originate the highest number of anomalous KPIs, considering the anomalous KPIs that are central to the propagation of the anomalies. This information can be useful to the operator to interpret a detected problem.

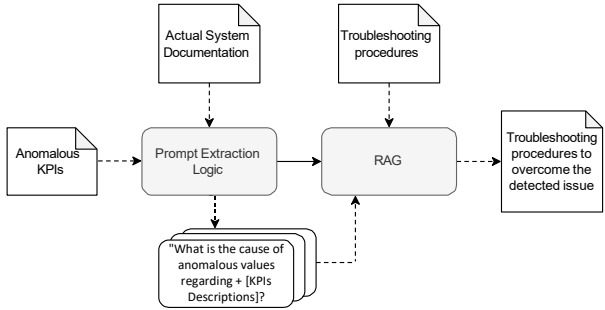

Fig. 4. Overview of REACT.

*B. React*

Figure 4 illustrates the two main components of the REACT system: *Prompt Extraction Logic* and *RAG*.

The *Prompt Extraction Logic* takes as input the anomalous KPIs along with their associated anomaly scores, as well as the *Actual System Documentation*, which is a mapping between KPIs and their definitions. This component exploits these inputs to automatically generate a prompt for querying the RAG, focusing on the identification of the troubleshooting procedures relevant to the detected issues. Specifically, it uses the top 2 to 4 KPIs with the highest anomaly scores to create the following prompt for the RAG: *"What is the cause of anomalous values regarding + [KPIs Descriptions]"*. This range of KPIs was selected after preliminary experiments, where using fewer KPIs often produced under-specified prompts, while using more introduced excessive noise.

Once the prompt is submitted to the *RAG*, the RAG accesses its knowledge base containing the system's troubleshooting procedures extracted directly from the technical manuals. The RAG involves an *LLM* which, upon receiving the prompt, retrieves the most relevant chunks of information based on semantic similarity between the question and the content of the knowledge base, augmenting the prompt with the retrieved information, specifically, a selection of the company's troubleshooting manuals.

For example, the RAG can automatically augment the prompt with all the reasons why the Air Compressed System may not start, if the original prompt is asking for the causes and troubleshooting procedures to be applied to an Air Compressed System that is off. The LLM can thus exploit this list of causes present in the prompt to formulate a useful answer, which could not consider the specific procedures available for the Air Compressed System otherwise.

The final output provided to the operator consists of *troubleshooting procedures to overcome the detected issue*. This offers actionable and targeted guidance to the operator. Moreover, the RAG provides links to the items of the knowledge base that have been used to produce the answer. The operator can access these portions of the manuals to verify the suggested procedure before actuating it.

For example, the RAG can suggest how to exactly check cables, connector terminators, switches, and other components when the observed problem affects electric equipment for which the manuals have troubleshooting procedures defined.

### III. EMPIRICAL ASSESSMENT

Our experimental evaluation focused on assessing both the *accuracy* of PREVENT and the *efficacy* of REACT in providing accurate troubleshooting guidance and retrieving relevant procedures.

*A. Accuracy of Prevent*

We applied PREVENT to an industrial case we have access to in the context of our collaboration with Fincantieri. To increase the generality of our results, we also included results from a publicly available dataset about the behavior of a CNC milling machine[3]. For both cases, we measured the number of false positives (FPs) and correct failure predictions, as the parameter *sigma* varied between 1.5 and 10.5.

*Fincantieri Case:* In the Fincantieri case study, we evaluated PREVENT on both failure-free and faulty scenarios. These scenarios represent various executions of an *Air Compressed System*. The actual system plays a critical role in ensuring the safe and efficient operation of ships. For this reason, predicting failures in this system is a crucial and complex task due also to the heterogeneity of the interacting components (e.g., Diesel Generators, Propulsion Diesel Engines).

We trained PREVENT on the provided failure-free simulation scenarios. Since PREVENT can be trained using failure-free scenarios only, we used the Elbow Method [5] to appreciate the degree of improvement in the number of false positives obtained across the four failure-free scenarios we had access to, when a higher *sigma* is used. Figure 5 shows the results.

The *elbow* point appears at *sigma* equals 4.5. However, to reduce the number of FPs across all scenarios, a more conservative threshold between 6 and 9 is advisable.

Table I presents the results for eight representative faulty scenarios provided by the company, using the same range of *sigma* values (here a false positive is an anomalous value detected before the fault is activated). The number of FPs drops significantly at *sigma* equal to 4.5, consistently with the evidence collected with failure-free scenarios. However, only for *sigma* values above 6, some scenarios show no FPs at all. Cells highlighted in red in Table I indicate cases where no failure were predicted, suggesting that overly conservative thresholds may lead to missed predictions.

---
[3]https://www.kaggle.com/datasets/shasun/tool-wear-detection-in-cnc-mill/data

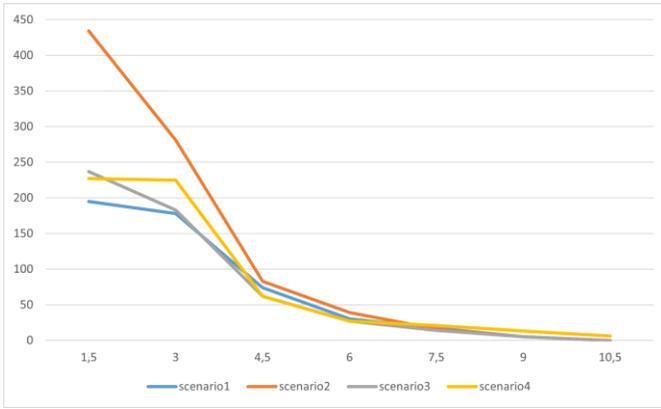

Fig. 5. Elbow Method applied with failure-free scenarios from Fincantieri.

| sigma | fault1 | fault2 | fault3 | fault4 | fault5 | fault6 | fault7 | fault8 |
|---|---|---|---|---|---|---|---|---|
| 1.5 | 222 | 135 | 236 | 215 | 633 | 406 | 242 | 396 |
| 3 | 116 | 97 | 193 | 206 | 265 | 338 | 149 | 281 |
| 4.5 | 35 | 46 | 40 | 59 | 34 | 39 | 5 | 25 |
| 6 | 10 | 20 | 14 | 27 | 0 | 0 | 0 | 0 |
| 7.5 | 3 | 8 | 7 | 11 | 0 | 0 | 0 | 0 |
| 9 | 0 | 2 | 1 | 10 | 0 | 0 | 0 | 0 |
| 10.5 | 0 | 0 | 0 | 9 | 0 | 0 | 0 | 0 |

TABLE I
NUMBERS OF FALSE POSITIVES IN THE FAULTY SCENARIOS.

In the considered scenario, a threshold between 7.5 and 9 provides a good balance between failure detection and a limited number of false alarms produced.

*CNC milling machine Case:* In the CNC case study, we also tested PREVENT on both the failure-free and faulty scenarios. Figure 6 shows the number of false positives detected in the failure-free scenarios for different values of *sigma*. Also in this case, the elbow consistently appears at *sigma* equals to 4.5.

For the faulty scenarios, Table II presents the number of failure predictions for each *sigma* value. Based on the documentation provided, we assume that each scenario exhibits faulty behavior throughout its entire duration. As with the failure-free cases, a more conservative choice of *sigma* (i.e., values closer to 6 or 9) provides a better tradeoff between true and false positives.

The experience with the Fincantieri's system, corroborated by the results obtained with the CNC case study, shows that the application of PREVENT can be successful, at the cost of setting an appropriate configuration of the parameters, which can be identified from failure-free executions by making a conservative choice based on the Elbow method. On the negative side, results are quite sensitive to the selection of this threshold, which may introduce too many false positives if not selected appropriately. Thus, the CNC case study not only strengthens the evidence PREVENT can generalize across different domains, but also confirms the results obtained with the Fincantieri's system.

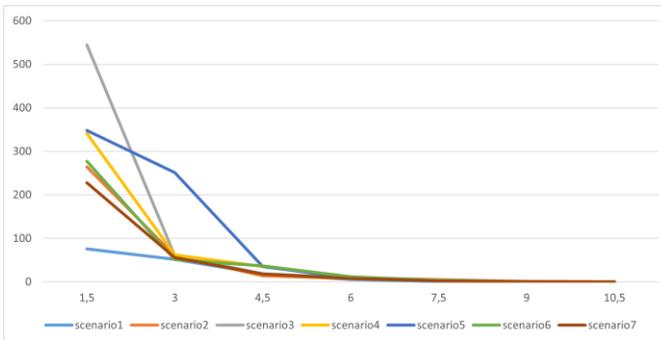

Fig. 6. Elbow Method applied to the failure-free scenarios in CNC.

| sigma | fault1 | fault2 | fault3 | fault4 | fault5 | fault6 | fault7 | fault8 | fault9 | fault10 |
|---|---|---|---|---|---|---|---|---|---|---|
| 1.5 | 409 | 407 | 166 | 260 | 233 | 198 | 552 | 210 | 273 | 471 |
| 3 | 192 | 76 | 56 | 89 | 48 | 47 | 58 | 55 | 56 | 77 |
| 4.5 | 44 | 45 | 20 | 30 | 15 | 32 | 18 | 27 | 27 | 54 |
| 6 | 20 | 19 | 8 | 19 | 8 | 11 | 6 | 10 | 7 | 21 |
| 7.5 | 13 | 10 | 1 | 14 | 1 | 3 | 2 | 4 | 1 | 13 |
| 9 | 5 | 2 | 0 | 4 | 0 | 1 | 1 | 0 | 0 | 7 |
| 10.5 | 0 | 1 | 0 | 1 | 0 | 0 | 0 | 0 | 0 | 2 |

TABLE II
NUMBERS OF FAILURE PREDICTIONS IN THE FAULTY SCENARIOS FOR THE CNC CASE STUDY.

### B. Effectiveness of React

To evaluate REACT, we used as input the anomalous KPIs detected in the eight faulty scenarios of the Fincantieri case study (we cannot apply REACT to the CNC milling machine since manuals with troubleshooting procedures are not available). The system documentation provided by Fincantieri's experts, alongside the simulation data, served as reference material. This documentation includes both a description of the simulation scenarios and a table listing each KPI monitored during operation, which we leveraged in prompt generation.

In our experimentation, the RAG architecture was built using *Ollama*[4] as the container, *LangChain*[5] as RAG framework, *Qwen2.5*[6] as LLM, *Chroma*[7] as Vector Store, *nomic-embed-text*[8] as embeddings component.

For each faulty scenario, after a failure is predicted by PREVENT, REACT generates a prompt with the top anomalous KPIs to retrieve the corresponding troubleshooting procedure. In our experiments with the faults provided by Fincantieri, REACT selected troubleshooting procedures about either the propulsion diesel engines or the air pressure of the tank. This result is quite promising since the selected procedures are distinct and specifically related to the simulation scenarios under analysis

To assess the quality of the responses generated by RE-ACT, we consulted the expert of the system available from Fincantieri, who reviewed and provided feedback about the output of REACT. The expert confirmed that REACT returned relevant troubleshooting procedures and that the responses were consistent with the injected faults. However, a specific observation was made regarding the scenarios involving the propulsion diesel engine failures. In these cases, the failures were preceded by a degradation in the compressor performance, evidenced by a significant increase in recharge time. The expert noted that the REACT's responses did not reference the compressor, which would have been a relevant resource to thoroughly check the condition of the system.

---

[4] https://ollama.com/
[5] https://www.langchain.com/
[6] https://huggingface.co/Qwen
[7] https://www.trychroma.com/
[8] https://ollama.com/library/nomic-embed-text

This feedback highlights an opportunity for improvement in enhancing the integration between anomalous operational data generated by PREVENT and the technical documentation retrieved by REACT. In particular, more accurate diagnostic processes might be able to analyze the point-to-point evolution of failures, rather than only retrieving troubleshooting procedures at failure prediction time. Future work will focus on mapping intermediate degradation patterns to the relevant documentation for more accurate troubleshooting support.

## IV. LESSON LEARNED

Based on the experience gained through the experiments presented in this paper, we identified some actionable insights:

**It is possible to close the gap between low-level anomaly detection data and high-level user-oriented troubleshooting procedures**: Anomaly detection systems refer to low-level metrics that can be captured from complex systems in operation. On the contrary, troubleshooting procedures are user-oriented manuals that describe the procedures that operators have to follow to overcome problems. The integration of PREVENT with the newly defined module REACT is *a first evidence of how LLMs, and RAGs in particular, can be used to close the gap between system-oriented evidence and user-oriented procedures*, and increase the level of automation in problem resolution.

**Unsupervised anomaly detection methods can be effective, if properly configured**: Our experience shows that unsupervised methods can be practical, as also reported in other studies [1], [7], [9], [17]. However, fine-tuning parameters relying on failure-free execution only might be challenging, especially when the operated software is a non-trivial industrial system. We observed, especially in the Air Compressed System, that applying overly conservative thresholds aimed at minimizing false positives can lead to the unintended consequence of missing actual failures. Relying solely on the Elbow Method may not yield satisfactory results, and setting appropriate parameters requires experts' knowledge. For example, Figure 5 shows that the elbow occurs at *sigma* equal to 4.5, yet the number of false positives is too high at that point, and a more conservative choice is necessary. This highlights the *importance of the expert's domain knowledge*, such as understanding the expected variability of operational signals and the acceptable level of noise in real executions, which allows selecting thresholds that better reflect realistic system behavior. This manual tuning that may limit full automation represents a lightweight source of intervention compared to traditional rule-based approaches and does not compromise the scalability of the method across different systems. We exploited the Elbow method since it is a widely established and used method. The possibility of investigating and comparing more sophisticated methods in subsequent studies to improve the robustness and effectiveness of the system is certainly a direction for future research.

**Sophisticated strategies are needed to systematically capture the evolution of failures**: The proposed strategy was able to capture the main symptoms of failures and recommend appropriate troubleshooting procedures, but Fincantieri's experts reported the lack of some intermediate events in the retrieved descriptions. This opens to more *research about how to exploit the data from anomaly detection to generate a better and more comprehensive description of the evolution of failures*, and allow operators to make more informed decisions.

**Importance of KPIs selection for prompt generation**: During the experiments with REACT, we noticed that the quality of the information retrieved with the prompts significantly impacted the relevance of the retrieved troubleshooting procedures. Initially, we built prompts with all the KPIs labeled as anomalous by PREVENT, resulting in overly broad queries involving many metrics. This made it difficult for the RAG to focus on a specific issue, and the retrieved documentation was often fragmented or unrelated. To address this, we introduced a filtering mechanism to select only the most relevant KPIs, focusing on the two to four most suspicious KPIs. This balance proved essential for generating meaningful and actionable responses. This result also suggests that *more work is needed on the strategies to generate prompts from anomalous KPIs*.

## V. RELATED WORK

Understanding and addressing failures in complex industrial systems requires a multifaceted approach that spans both automated diagnostics and human-in-the-loop support. In this section, we discuss prior work in two key areas: (i) failure prediction and fault localization, where the focus is on automatic detection and identification of root causes of system anomalies, and (ii) troubleshooting support, which aims to assist human operators in resolving faults efficiently.

*Failure Prediction and Fault Localization.* Failure prediction and fault localization in industrial systems have been extensively studied using a variety of approaches. Traditional methods include rule-based systems [4], signature-based techniques [8], [10], [19], and semi-supervised learning [13], [20], relying on prior knowledge or labeled failure data. These assumptions often limit their applicability in real-world industrial contexts, where such data may be scarce or unavailable.

Recent work has explored unsupervised learning techniques that learn normal system behavior and detect deviations without requiring labeled failure data [1], [7], [9], [17]. However, many of these approaches focus solely on anomaly detection, offering limited support for fault localization.

To address this, recent approaches have combined anomaly detection with causality analysis [6], [12], [15], [21]. This enables a more interpretable and structured diagnosis process, especially in distributed and complex systems where failures propagate across multiple components.

Our approach uses PREVENT that offers a fully unsupervised solution that not only detects anomalies but also localizes faults using causality analysis. Moreover, we apply it to naval systems, a domain characterized by complex interactions between mechanical, software, and hardware components, which remains underexplored in the literature.

*Troubleshooting and Human-in-the-Loop Support.* While failure prediction is critical, many industrial systems still require human intervention to resolve faults. In such cases, effective troubleshooting support becomes essential. Recent advances have taken advantage of natural language processing and information retrieval to improve maintenance workflows [2], [16], [18].

Ren et al. [16] developed a voice-interactive fault diagnosis system for industrial robots, enabling operators to retrieve information from manuals using voice commands. Kiangala and Wang [11] introduced a hybrid AI chatbot that combines generative and rule-based AI to support troubleshooting in Industry 5.0 environments. Su et al. [18] proposed an ontology-based method for aerospace maintenance, integrating information flow modeling and ant colony optimization to improve diagnosis and troubleshooting.

Alfeo et al. [2] highlighted the value of deep learning in retrieving solutions from historical technical assistance reports, demonstrating the potential of data-driven approaches in predictive maintenance.

Our work complements these efforts by introducing REACT, a RAG-based system that automatically generates prompts from anomaly data and retrieves relevant troubleshooting procedures from system documentation, exploring the possibility of closing the gap between the low-level anomaly detection data and the user-oriented documentation that are defined using largely disjoint terms. Unlike ontology-based or voice-driven systems, REACT leverages LLMs to interpret unstructured queries, enabling intuitive and context-aware interaction. This approach supports human-in-the-loop workflows in scenarios where automatic recovery is not feasible, bridging the gap between anomaly detection and actionable resolution.

## VI. Conclusions and Future Work

Large industrial systems often exhibit unpredictable behavior, making failures difficult to anticipate and overcome due to their inherent complexity. Traditional rule-based and signature-based approaches are constrained by the need for prior knowledge of typical failure patterns, information that is often unavailable or incomplete in real-world industrial settings. To address these limitations, we adopt an unsupervised methodology based on PREVENT, which combines anomaly detection with fault localization, extended with REACT, a novel module that can automatically retrieve troubleshooting procedures from a set of anomalous KPIs using a RAG.

Our findings show how anomaly detection and RAG can be used jointly to suggest troubleshooting procedures that can be actuated by users. At the same time, we discuss a lesson learned and open challenges to make these solutions more practical and effective.

Future work concerns expanding the evaluation to a broader range of scenarios, improving the alignment between anomaly descriptions and procedural documentation, and exploring the integration of interactive capabilities to better support human operators in the decision-making loop.

**Acknowledgements**. This work has been supported by ATOS project, funded by the MUR under the PNRR- CN - HPC - ICSC program (CUP: H43C22000520001).